\documentclass[twocolumn,showpacs,preprintnumbers,amsmath,amssymb]{revtex4}
\usepackage{graphicx}
\usepackage{subfigure}
\usepackage{verbatim}
\newcommand{\be}{\begin{eqnarray}}
\newcommand{\ee}{\end{eqnarray}}

\begin{document}

\title{Random Diffusion Model with Structure Corrections}
\author{David D. McCowan}
\author{Gene F. Mazenko}
\address{The James Franck Institute and the Department of Physics\\
The University of Chicago\\
Chicago, Illinois 60637}
\date{April 4, 2010}

\begin{abstract}

The random diffusion model is a continuum model for a conserved scalar density field $\phi$ driven by diffusive dynamics where the bare diffusion coefficient is density dependent. We generalize the model from one with a sharp wavenumber cutoff to one with a more natural large-wavenumber cutoff. We investigate whether the features seen previously -- namely a slowing down of the system and the development of a prepeak in the dynamic structure factor at a wavenumber below the first structure peak -- survive in this model. A method for extracting information about a hidden prepeak in experimental data is presented.

\end{abstract}

\pacs{05.70.Ln, 64.60.Cn, 64.60.My, 64.75.+g}

\maketitle

\section{Introduction}

In this work we study a generalization of the random diffusion model (RDM) introduced by Mazenko in Ref. \onlinecite{RDMI}, henceforth referred to as RDMI. The RDM is a model for nonlinear diffusion in colloidal systems. It is formulated in terms of a conserved density field and, in its simplest form, can be chosen to have Gaussian static statistics. The model can be motivated as a continuum generalization of the facilitated spin models of glassy dynamics \cite{FA1,FA2,chan1,chan2,WBG1,WBG2,jack}. These models have a density dependent kinetic coefficient which leads to slowing down in dense systems. The RDM similarly has a density dependent bare diffusion coefficient. In RDMI, this bare diffusion coefficient led to a slowing down as the density was increased. The more realistic RDM presented here exhibits similar behavior.

There has been much speculation but relatively few solid results in establishing the existence of a mode-coupling theory (MCT) ergodic-nonergodic (ENE) transition in the field-theoretic models of the liquid-glass transition. The RDM is a candidate for the simplest such model, however this point will not be a focus of this paper. In RDMI, the model was shown to undergo an ENE transition at one-loop order, but not at two-loop order. The RDM is part of a larger class of models with density dependent bare diffusion coefficients including those discussed by Dean \cite{Dean}, Kawasaki and Miyazima \cite{kaw}, and Miyasaki and Reichman \cite{RFT}.

In RDMI, the Fokker-Planck formalism was used to generate a self-consistent perturbation expansion for the memory function. This memory function could be inserted into the kinetic equation which describes the time-evolution of the dynamic correlation function, the physical observable of interest. In the simplest realization of this model, one has a single control parameter, $g$, which controls the density dependence. RDMI treated a course-grained system in what was termed the {\em structureless approximation}; the static structure factor of the system was assumed constant up to a wavenumber cutoff $\Lambda$.

Solving the kinetic equation revealed the following:

\begin{enumerate}
\item
As $g$ increases, the system slows down.

\item
A new peak, termed the prepeak, develops at a wavenumber $q_0$ away from zero. The form of this peak can be fit to a Gaussian
\be
f(q,t) = Ae^{-B(q-q_0)^2}
\ee
where the amplitude A decreases with time and the peak width $1/{\sqrt B}$ narrows with time. This form reveals a new growing length in the problem  (${\sqrt B}$) and fits of the the amplitude change smoothly from an exponential form at zero coupling to a power law form near a critical value.

\item
Above this critical value, the system becomes unstable and the prepeak amplitude grows without bound. Such behavior is clearly unphysical and as such, the model breaks down.
\end{enumerate}

In this paper, we generalize the random diffusion model to a more realistic form where wavenumber integrations are cut off naturally in the theory. This requires incorporating a more realistic static structure factor into the theory. Here, the integrals are naturally cut off by the large wavenumber decay of the direct correlation function. Our new model depends on two parameters; we have a coupling constant analogous to that seen in the structureless case, but our model also depends explicitly on density.

For this new model we find the following:

\begin{enumerate}
\item
As the coupling and density increase, the system again slows down.

\item
A prepeak forms, now located between $q=0$ and the first structure peak. This peak and the peaks due to the static structure factor can be fit to Gaussians with decaying amplitude and narrowing width just as before and the amplitude again changes from an exponential to a power law form as the system approaches the critical crossover.

\item
Above the critical values of density and coupling constant, we again find that the prepeak grows without bound. Whereas previously this separation between stable and unstable growth was characterized by a single number, our model has a separation given by a critical line in coupling constant-density parameter space.
\end{enumerate}

This improved model therefore preserves the features seen in RDMI.

In this work, we again discuss in detail the slowing down of the system and the interesting feature of the dynamically generated prepeak. Though the development of this prepeak in both RDMI and this paper is surprising, it is worth pursuing. The random diffusion model is simple, yet incorporates nonlinearities known to be present and the perturbation is straightforward and without the "tricks" sometimes used to force things into mode-coupling form. It is dynamically generated instead of arising via the static structure factor and we find here that it is robust in the choice of said statics; what one may have worried was an artifact of a coarse approximation in RDMI is shown here to survive in the same form.

The final feature of the model, the unstable growth of the system above critical values of the model parameters, is a feature which reveals the breakdown of the model. In this region, the system likely wants to nucleate, but no terms are present in the model to provide this stabilization. We discuss some loose features of the instability and then propose the terms, motivated by density function theory, which may cutoff the growth.

Because our model uses a realistic structure factor, we can draw a closer connection to experimental results than was possible in RDMI. In the last section of the paper we present a method by which one can extract information about a possible prepeak hidden in experimental measurements of the dynamic structure factor. Though few, some experimental allusions to prepeaks exist and comparison is desired.

\section{The Random Diffusion Model}
\subsection{Introduction}

Let us introduce the random diffusion model in the Fokker-Planck context. For a fundamental density field $\phi({\bf x})$, we take an effective Hamiltonian quadratic in $\phi$ given by
\be
{\cal H}_{\phi}=\frac{1}{2} \int d^d x_1 d^d x_2 \delta \phi({\bf x_1}) \chi^{-1}({\bf x_1}-{\bf x_2})
\delta \phi({\bf x_2})
\label{eq:Hamiltonian}
\ee
where $\delta\phi({\bf x})=\phi({\bf x})-n$ and $n=\langle\phi\rangle$. We will study density time correlations through the equilibrium intermediate dynamic structure factor given by
\be
C({\bf q_1},{\bf q_2};t)
&=&\langle\phi({\bf q_2},t)\phi({\bf q_1},0)\rangle
=\langle\phi({\bf q_2})e^{-\tilde{D}_{\phi}t}\phi({\bf q_1})\rangle
\nonumber\\
&=&(2\pi)^d \delta({\bf q_1}+{\bf q_2}) C({\bf q_1};t)
\ee
where $\phi({\bf q})$ is the Fourier transform of the fundamental field $\delta\phi$, averages are given by $\langle f(\phi) \rangle =\int {\cal D}(\phi)W_{\phi} f(\phi)$ with an equilibrium probability distribution of the usual form, $W_{\phi}=e^{-\beta\cal{H}_{\phi}}/Z$, and $\tilde{D}_{\phi}$ is the adjoint Fokker-Planck operator defined below. We will also make frequent use of the static (equal-time) correlation function given by
\be
\tilde{C}({\bf q_1},{\bf q_2})=C({\bf q_1},{\bf q_2};t=0)=(2\pi)^d \delta({\bf q_1}+{\bf q_2}) \tilde{C}({\bf q_1})
\ee
which is related to the static susceptibility appearing in Eq. \eqref{eq:Hamiltonian} by
\be
\tilde{C}({\bf q_1})=\beta^{-1}\chi({\bf q_1}).
\ee

The adjoint Fokker-Planck operator takes the form
\be
\tilde{D}_{\phi}=\int d^d x \int d^d y \bigg[\frac{\delta\cal{H}_{\phi}}{\delta\phi({\bf x})}
-k_bT\frac{\delta}{\delta\phi({\bf x})}\bigg]\Gamma_{\phi}({\bf x},{\bf y})\frac{\delta}{\delta\phi({\bf y})}
\ee
where $\Gamma_{\phi}$ is a transport matrix which incorporates the bare diffusion coefficient \cite{TM}. In RDMI, $\Gamma_{\phi}$ was given by
\be
\Gamma_{\phi}({\bf x},{\bf y})={\bf \nabla}_{x}{\bf \nabla}_{y}[D(\phi)\delta({\bf x}-{\bf y})]
\ee
where the bare diffusion coefficient $D(\phi)$ was
\be
D(\phi)=D_0+D_1\phi({\bf x})={\bar D}+D_1\delta\phi({\bf x})
\ee
with ${\bar D} = \langle D(\phi)\rangle = D_0+D_1 n$.
For this work, however, we choose a more general form
\be
\Gamma_{\phi}({\bf x},{\bf y})&=&{\bf \nabla}_{x} {\bf \nabla}_{y}\int d^dz\Big(f_0(x-z){\bar D}f_0(y-z)
\nonumber\\
&+&f_1(x-z)D_1\delta\phi(z)f_1(y-z)+... \Big).
\label{eq:TM}
\ee
By relaxing the constraining delta-function and introducing the general functions $f_0$ and $f_1$, we will have considerably more freedom in regulating integrals which result from application of perturbation theory. For now, we leave the functions undetermined, but will later set them in such a way as to give the short-time sum rules correctly and give the large-wavenumber dependence of the vertex in the memory function as in mode-coupling theory.

Our choice for the transport matrix (both the form of RDMI and the more general form here) is motivated as a course grained alternative to the microscopic form $D(\phi)=D_0$ most often used. Density dependent terms are expected to play a role in the dynamics and as such, have been incorporated here.

Our model therefore depends on the functions $f_{0}$, $f_{1}$ and $\chi$ and the constants
$D_{0}$ and $D_{1}$. The effects of extending the transport matrix to higher order through additional constants ($D_2$, etc.) and functions ($f_2$, etc.) will be addressed in a future paper.

The physical diffusion coefficient is given by
\be
D_p=\lim_{q\rightarrow 0} D(q)
\ee
where D(q) comes from the Fourier transform of the average of the transport matrix:
\be
D(q)q^2=\int d^d x_1 e^{i{\bf q}\cdot({\bf x_1}-{\bf x_2})}\langle\Gamma_{\phi}({\bf x_1}-{\bf x_2})\rangle.
\ee
Thus, we have
\be
D_p= {\bar D} f^2_0(0)
\ee
where we have used the fact that averages over odd powers of $\delta\phi$ vanish.

\subsection{Memory Function Formalism}

In order to study the time evolution of the dynamic structure factor $C(q,t)$, we organize our theory using the memory function technique \cite{NESM}. In Laplace transformed space, the kinetic equation is given by
\be
[z+K(q,z)]C(q,z)={\tilde C}(q)
\ee
where
\be
C(q,z)=-i\int^{\infty}_0 dt e^{izt} C(q,t)
\ee
and where $K({\bf q},t)=K^{(s)}({\bf q})+K^{(d)}({\bf q},t)$ is the memory function decomposed into a static piece and a dynamic piece. Inverse Laplace transforming this equation gives the form
\be
\frac{\partial C(q,t)}{\partial t} &=& iK^{(s)}(q)C(q,t)
\nonumber\\
&+&\int^t_0 ds K^{(d)}(q,t-s) C(q,s).
\label{eq:KE}
\ee

First, the static memory function is determined (without approximation) by the equilibrium average
\be
K^{(s)}(q){\tilde C}(q) = i\beta^{-1}\langle\Gamma_{\phi}(q)\rangle.
\ee
Using our transport matrix given by Eq. \eqref{eq:TM}, we have
\be
K^{(s)}({\bf q}) &=& i\beta^{-1}q^2 {\bar D} f_0^2(q)\tilde{C}^{-1}(q)
\nonumber\\
&=& iq^2 {\bar D} f_0^2(q)\chi^{-1}(q).
\ee

The dynamic part of the memory function is more complicated to compute and can be determined via a perturbation expansion. If we assume that the nonlinearities are small, we may expand in powers of $D_1$, the coefficient of the density correction to the diffusion coefficient. A full derivation is given in RDMI, but at lowest order, the dynamic memory function is given by
\be
&&K^{(d)}({\bf q},t)\tilde{C}(q)
\nonumber\\
&&=-2\int \frac{d^dk}{(2\pi)^d}[V({\bf q},{\bf k})]^2
C({\bf k};t)C({\bf q}-{\bf k};t)
\label{eq:dynamic}
\ee
where $V({\bf q},{\bf k})$ is the vertex function. We define our vertex in light of our new choice for $\Gamma_{\phi}$ as
\be
V({\bf q},{\bf k})&=&\frac{i}{2}D_1[f_1(q){\bf q}
\cdot f_1(k){\bf k}\chi^{-1}(k)
\nonumber\\
&+& f_1(q){\bf q}\cdot f_1({\bf q}-{\bf k})({\bf q}-{\bf k})\chi^{-1}({\bf q}-{\bf k})]
\ee
which has a simple dependence on the function $f_1$. We choose this form to mimic the traditional mode-coupling theory form. (For more on traditional MCT, see, e.g. Refs. \onlinecite{Das} and \onlinecite{Goetze}.) Now that we have defined this vertex, it is set to arbitrarily high order.

\subsection{Structure corrections}

Perturbation expansions inevitably run into wavenumber integrals such as the one we derived for the dynamic memory function, Eq. \eqref{eq:dynamic}, which are potentially divergent. Care must be taken when forming and evaluating the integrals, and one usually must either develop a rationale for implementing a large-wavenumber cutoff or structure the vertex such that the integrals remain finite.

Constructing a vertex which enforces convergence can be difficult. For example, Ref. \cite{ABL} studies the Dean-Kawasaki model \cite{Dean, KK} applicable to colloids. This work analyzes the theory in terms of time-reversal symmetry, but this approach ultimately leads to a vertex that is not of a standard MCT form and to divergent integrals. They propose that the solution may be a cumbersome resummation of higher-order diagrams to renormalize the vertex, but leave a full solution to future work.

For the random diffusion model, divergent integrals first appeared in RDMI. The approach in that work was to take a sharp cutoff at finite wavenumber $\Lambda$ and (for most of the paper) to take the static susceptibility to be constant such that $\chi^{-1}(q)=r$. This approach was called the {\it structureless approximation} because it corresponded to a coarse-grained system where the short-distance degrees of freedom (including the first peak in the static structure factor) had been integrated out \cite{cutoff}. While this simplified the problem and kept the integrals finite, one has reason to worry whether the results are strongly influenced by the nature of the cutoff and the {\it structureless} nature of the structure factor. It is therefore desirable to to impose a different method to regulate the integrals which does not raise such concerns.

To regulate the wavenumber integrals in this work, we now define our functions $\chi$, $f_0$, $f_1$ introduced earlier. First, let us restore the {\it structure} to the model by defining $\chi(q)$ in terms of the full static structure factor, $S(q)$, as
\be
\chi(q) = n\beta S(q)
\ee
where
\be
S(q) = \frac{1}{1-nC_D(q)}
\ee
and $C_D(q)$ is the direct correlation function which decays to zero at large wavenumber. For simplicity, we will use the solution to the Percus-Yevick approximation for hard-spheres in this paper. (See e.g. Refs. \onlinecite{ashcroft} and \onlinecite{Hansen} for details.) In principle, one could take any realistic analytic approximation for $S(q)$ or use experimental results.

Next, the function $f_0(q)$ appears in the static memory function $K^{(s)}(q)$ and the physical diffusion coefficient $D_p(q)$. Since we have no regularization constraints on either of these functions, let us take $f_0(q)=1$ which gives
\be
K^{(s)}(q)=\frac{iq^2 {\bar D} f_0^2(q)}{\chi(q)}=\frac{iq^2 {\bar D}}{\tilde{C}(q)\beta}
\label{eq:static}
\ee
and
\be
D_p={\bar D} f_0^2(0) = {\bar D}.
\ee

Finally, the function $f_1(q)$ appears only in the vertex (and therefore in the dynamic memory function $K^{(d)}(q,t)$). As discussed above, we want to choose $f_1(q)$ so that the memory function integral remains finite as we take the cutoff to infinity. Let us reason our way through the appropriate choice.

As it stands, the long wavenumber behavior of the vertex (beside the explicit $q^2$ dependence) is governed by $\chi^{-1}(q)$ which approaches unity at as $q\rightarrow\infty$. Let us instead let the vertex go as the direct correlation function $C_D(q)$ which approaches zero in the same limit. Thus, we want $f_1(q)\sim \chi(q) C_D(q)$. Normalizing $f_1(q)$ to unity at $q=0$, we therefore have
\be
f_1(q)=\frac{\chi(q) C_D(q)}{\chi(0) C_D(0)}=\frac{S(q)nC_D(q)}{S(0)nC_D(0)}.
\ee
Note that if we now put $f_1(q)$ back into the bare diffusion coefficient (Eq. \eqref{eq:TM}) and take the large-$q$ (short distance) limit, the density dependent $D_1$ term goes to zero returning the microscopic result $D(\phi)=\bar{D}$.

This now gives the vertex
\be
V(q,k)&=&\frac{iD_1}{2n\beta} \frac{S(q)nC_D(q)}{(S(0)nC_D(0))^2}
\nonumber\\
&\times& {\bf q}\cdot[{\bf k}nC_D({\bf k})+({\bf q}-{\bf k})nC_D({\bf q}-{\bf k})].
\ee
This is the traditional MCT form which decays to zero as either $q$ or $k\rightarrow\infty$. Such behavior, we will see, is sufficient to keep the memory function finite without imposing a finite integral cutoff.

To simplify the kinetic equation let us now introduce dimensionless variables. The dimensionless wavenumber, time and density (packing fraction) are given by
\be
Q=q\sigma,
\ee
\be
T={\bar D} \sigma t \beta^{-1}
\ee
and
\be
\eta = (\pi/6) n\sigma^3
\ee
where $\sigma$ is the hard sphere diameter and the dimensionless correlation function is given by
\be
f(Q,T)=C(Q,T)/\tilde{C}(Q).
\ee

After inserting the static and kinetic equations in terms of dimensionless variables, we have
\be
\frac {\partial f(Q,T)}{\partial T} &=& -\frac{Q^2}{S(Q)} \frac{\pi}{6\eta}f(Q,T)
\nonumber\\
&&+R^2\int^T_0 dS N(Q,T-S)f(Q,S)
\label{eq:final}
\ee
where we've defined the simplified memory function
\begin{widetext}
\be
N(Q,T)&=&\frac{1}{2} \bigg(\frac{\pi}{6\eta}\bigg)^3 \frac{S(Q)n^2C_D^2(Q)}{S^4(0)n^4C_D^4(0)}
\int \frac{d^3K}{(2\pi)^3} \bigg[{\bf Q}\cdot{\bf K}nC_D(K)+{\bf Q}\cdot({\bf Q}-{\bf K})nC_D({\bf Q}-{\bf K})\bigg]^2
\nonumber\\
&\times& S(K)f(K,T)S({\bf Q}-{\bf K})f({\bf Q}-{\bf K},T)
\label{eq:kernel}
\ee
\end{widetext}
and the coupling constant
\be
R=\frac{D_1n}{{\bar D}}=\frac{D_1n}{D_0+nD_1}.
\ee

Let us make a few comments on this form.

First, the kinetic equation now depends only on two parameters, the density $\eta$ and the coupling constant $R$. Note that the temperature dependence has been absorbed into the dimensionless variables. The kinetic equation can be solved numerically to get $f(Q,T)$ and the general behavior as $\eta$ and $R$ are varied can be studied.

Second, the coupling constant $R$ can be either positive or negative (as the coefficient $D_1$ can be either positive or negative). However, since $R$ enters only through $R^2$, the overall sign of $D_1$ is irrelevant. We choose for our purposes the constraint $0\leq R\leq 1$. One can imagine combinations involving negative, but large values of $D_1n$ such that $|R|>1$, but since our perturbation expansion assumed small $D_1$, we are already overly generous by studying $R$ up to unity.

Finally, note that our form for $\partial f(Q,T)/\partial T$ diverges as $\eta^{-1}$ as $\eta\rightarrow 0$. While this may seem strange, this is just a manifestation of the characteristic time scale of the system slowing down with density. One can choose to rescale the dimensionless time as $T\rightarrow T^{\prime} = T\pi/6\eta$ and the explicit divergence disappears. For various reasons, we have chosen to use the unscaled time, though if one were interested in low density systems, the scaled variable would be a more appropriate choice.

\section{Numerical Results}
\subsection{RDMI: A Review}
Before we solve our model, let us review in more detail the results from RDMI. In that model, the kinetic equation (at lowest order) is given by
\be
\frac {\partial f(Q,T)}{\partial T} &=& -Q^2 f(Q,T)
\nonumber\\
&+&Q^4\int^T_0 dS N(Q,T-S)f(Q,S)
\ee
with
\be
N(Q,T)=g \int \frac{d^3K}{(2\pi)^3}f(K,T)f({\bf Q}-{\bf K},T)
\ee
and
\be
g=\frac{1}{2} \bigg(\frac{D_1}{{\bar D}}\bigg)^2 {\tilde C} \Lambda^3
\ee
where ${\tilde C}$ is a constant and $\Lambda$ is the wavenumber cutoff.

In the non-interacting case ($g=0$), the equation can be solved analytically to get the simple form
\be
f_0(Q,T)=e^{-Q^2T}.
\ee

As the coupling $g$ increases, the decay at large wavenumber values slows and a small peak develops. This peak (termed the prepeak) can be fit to a Gaussian of the form
\be
f_p(Q,T)=Ae^{-B(Q-Q_0)^2}
\ee
where the amplitude $A$ has the time dependence
\be
A(T) = A_0 \frac{e^{-ET}}{(T+T_A)^{\alpha}}
\ee
while $B$ satisfies
\be
B(T) = B_0 (T+T_B)^{\beta}
\ee
where $Q_0$, $A_0$, $B_0$, $T_A$, $T_B$, $\alpha$, $\beta$ and $E$ are positive, time-independent fit parameters.

\begin{figure}[btp]
\centering
\includegraphics[width=\columnwidth]{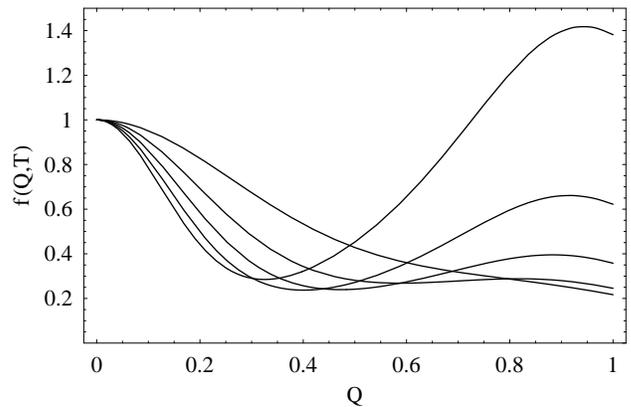}
\caption{A plot of $f(Q,T)$ as determined in RDMI for a coupling $g>g^*$. Each line is equally spaced in time and the times are earliest to latest as the peak amplitude goes from smallest to largest. Wavenumber space is normalized such that the cutoff corresponds with $Q$=1. $f(Q,T)$=0 for larger $Q$.}
\label{fig:RDMI}
\end{figure}

As the system approaches the critical coupling $g^*$, the peak amplitude goes to a completely algebraic decay with time such that $E\rightarrow0$. Above $g^*$, the above analytic equations cease to hold; the peak no longer decays at long times, but instead grows without bound, rendering the system unstable. Examples of the behavior of this model can be seen in Fig. \ref{fig:RDMI}.

\subsection{General Results}
For our model, the evolution of the dynamic correlation function is given by Eq. \eqref{eq:final}. While similar to the form of RDMI, the incorporation of a realistic static structure factor leads to a slightly more complicated form. We can numerically solve the kinetic equation for a particular set of the parameters $\eta$ and $R$. (The details of this solution are given in Appendix A.) The upper bound on the value of $\eta$ is $0.74$ which is the packing fraction of a close-packed solid. The upper limit for $R$ is $1$ as discussed at the end of the last section.

Before looking at the complete problem, let us first examine the simpler $R=0$ case to see some of the features which are common to all the solutions of the kinetic equation. In this limit there is no dynamic feedback from the memory function and the kinetic equation can be solved analytically to give
\be
f(Q,T)=exp\bigg(\frac{-Q^2T\pi}{6\eta S(Q)}\bigg).
\label{eq:R0}
\ee
Thus, the correlation function decays exponentially with time and its wavenumber dependence is strongly determined by the static structure factor $S(Q)$. This is the well known de Gennes narrowing form if we take a scaled time $T^{\prime}=T\pi/6\eta$.

\begin{figure}[btp]
\centering
\subfigure[]{\includegraphics[width=.45\columnwidth]{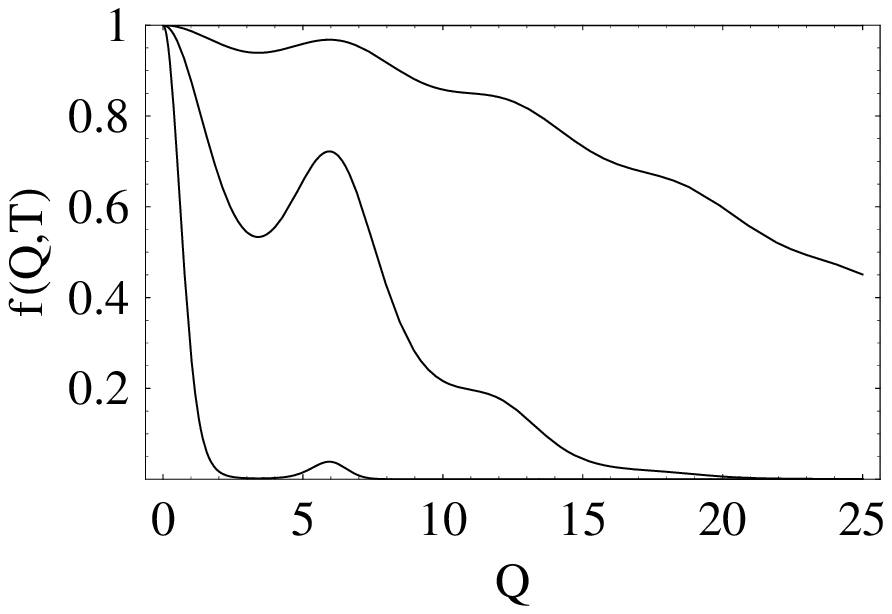}}
\subfigure[]{\includegraphics[width=.45\columnwidth]{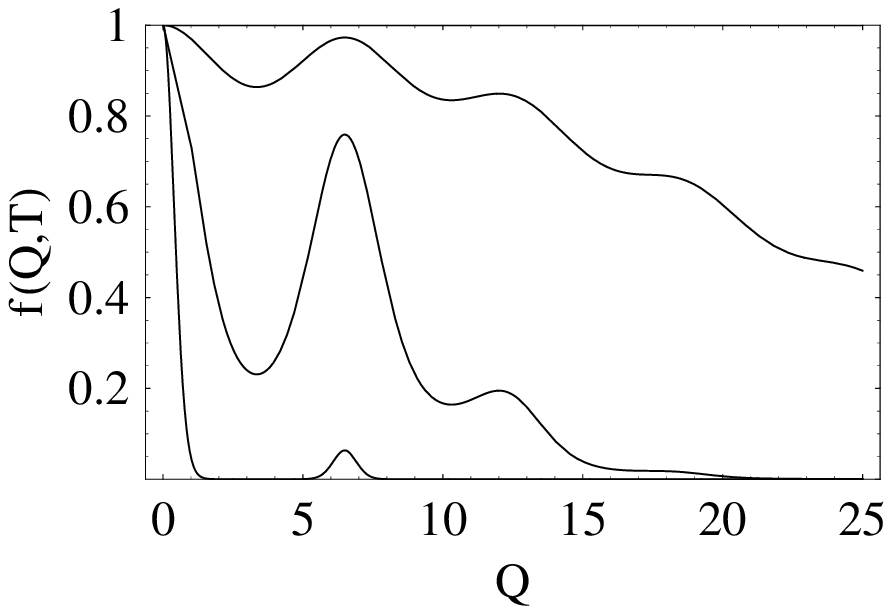}}
\subfigure[]{\includegraphics[width=.45\columnwidth]{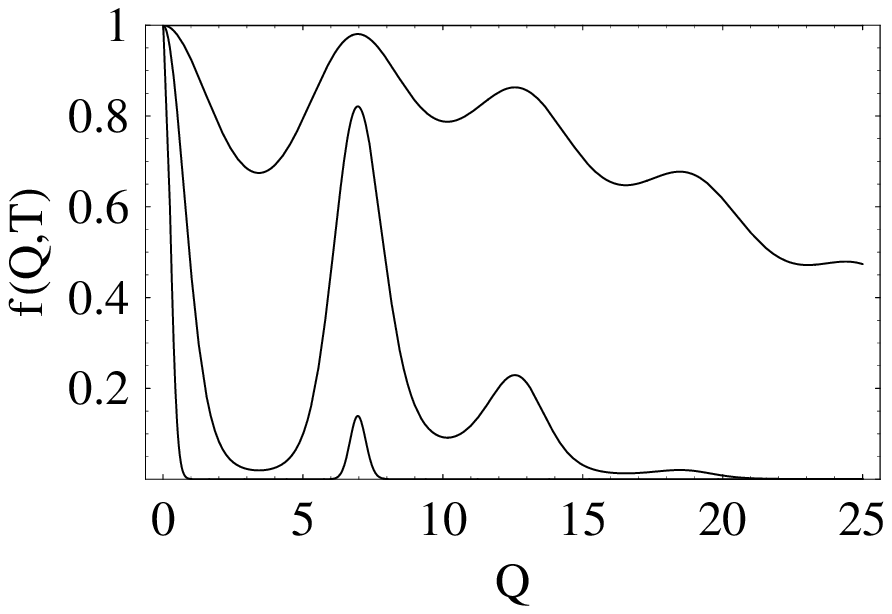}}
\subfigure[]{\includegraphics[width=.45\columnwidth]{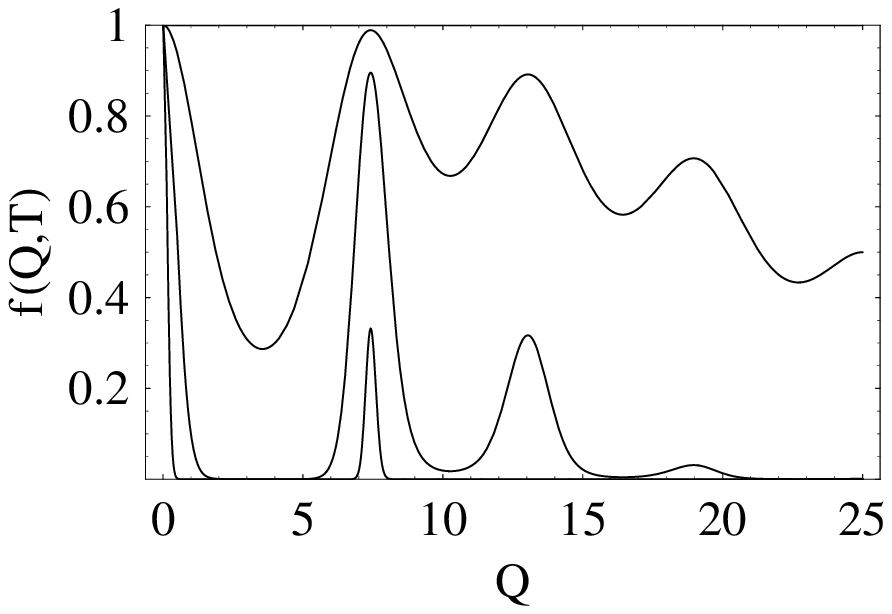}}
\caption{Plots of $f(Q,T)$ with $R=0$ for (a) $\eta=0.3$, (b) $\eta=0.4$, (c) $\eta=0.5$ and (d) $\eta=0.6$. Times are (from top line to bottom line) $T$=0.001, 0.01, and 0.1.}\label{fig:R0}
\end{figure}

In Fig. \ref{fig:R0} we plot the behavior of $f(Q,T)$ for various values of $\eta$. We see that a number of peaks form which decay to zero with time. Beside the peak centered at $Q$=0, these peaks are centered at the same wavenumbers as the peaks of the static structure factor. Even with $R=0$, we see that we have considerable structure in the theory.

We now turn to the case of $R\neq0$.

\begin{figure}[btp]
\centering
\subfigure[]{\includegraphics[width=\columnwidth]{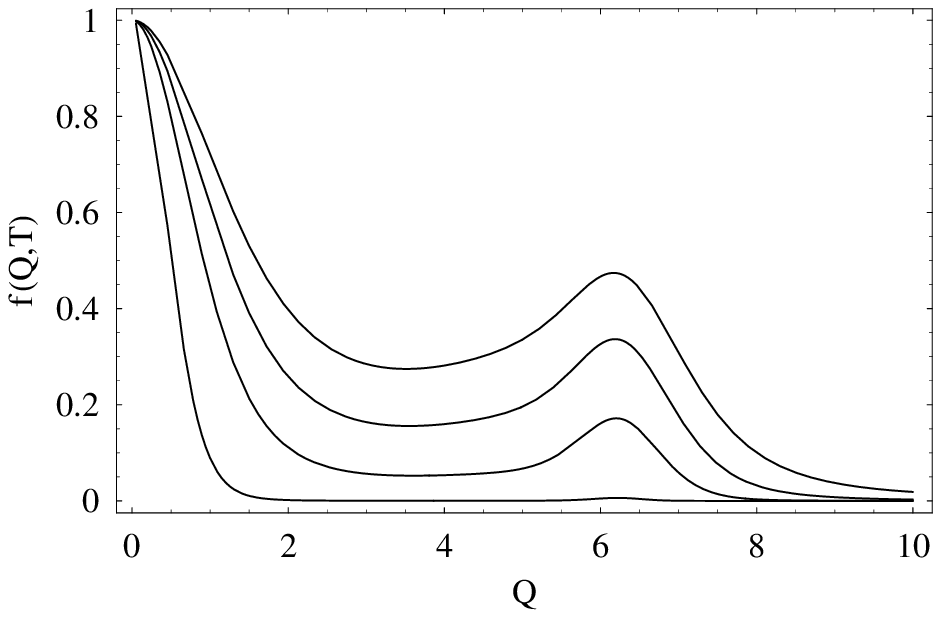}\label{fig:eta30a}}
\subfigure[]{\includegraphics[width=\columnwidth]{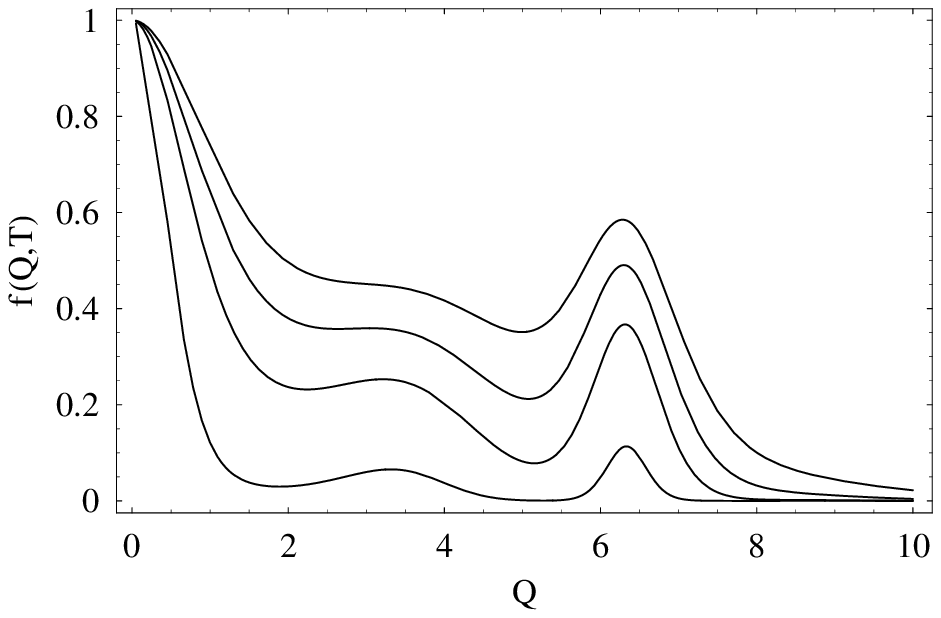}\label{fig:eta30b}}
\caption{Plots of $f(Q,T)$ for $\eta=0.30$ with (a) $R=0.51$ and (b) $R=1$. Times are (from top to bottom) $T$=0.02, 0.03, 0.05 and 0.15. Note that a prepeak is developing in the $R=1$ plot, however both plots decay to zero with time.}
\label{fig:eta30}
\end{figure}

At small $\eta$, say $\eta=0.10$ or $0.20$, the behavior is nearly indistinguishable from the $R=0$ case even as $R$ is increased through all allowed values. As we increase to intermediate values of $\eta$, say $\eta=0.30$, the situation becomes more interesting. In Fig. \ref{fig:eta30a} we plot results for $\eta=0.30$ and $R=0.51$ and see a slight slowing down of the decay. That is, the peaks in the $R=0.51$ case persist longer before decaying to zero than those in the $R=0$ case.

If we stay at these intermediate densities and ratchet up to extremely large $R$, we see an interesting new feature; between the Q=0 peak and the first structure peak, a new small peak appears. With time, this peak also decays away to zero and disappears before the first structure peak. Because of its position in Q-space, we will call this peak the {\em prepeak} because it is located at a smaller Q value than the first structure peak. As we push the coupling up to the limit $R=1$, we find that the prepeak persists for longer and longer times, but still decays to zero with all the other peaks. (Fig. \ref{fig:eta30b}.)

Looking at the behavior of the second and higher structure peaks, we find little of interest. These higher order peaks mirror the behavior of the first structure peak -- their decay slows with increasing R -- and we find no new features (e.g., higher order prepeaks). This remains true for most of the work presented in this paper and we will therefore restrict ourselves to discussion of the behavior at wavenumbers around and below the first structure peak unless otherwise noted.

\begin{figure}[btp]
\centering
\subfigure[]{\includegraphics[width=\columnwidth]{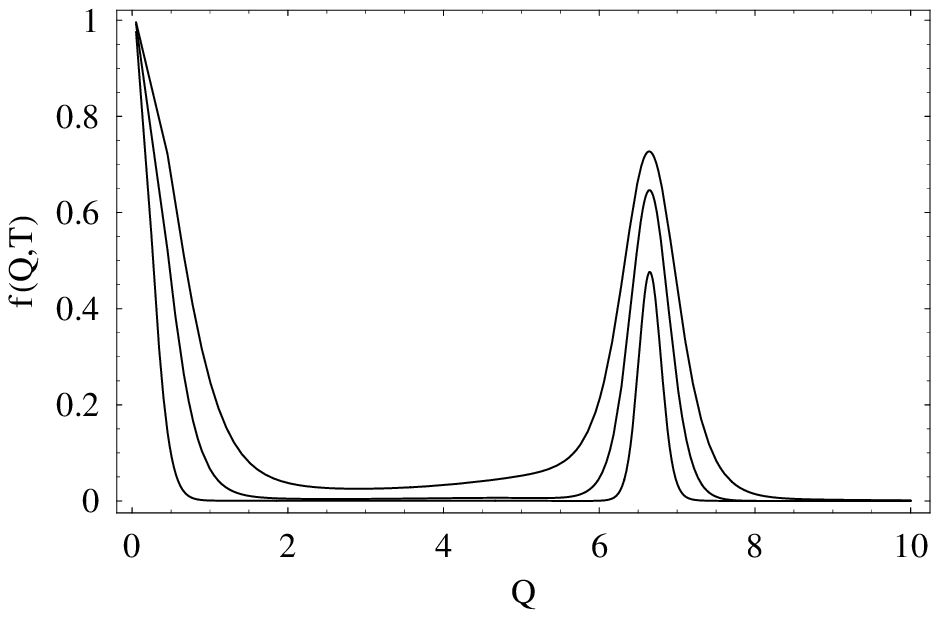}\label{fig:eta40a}}
\subfigure[]{\includegraphics[width=\columnwidth]{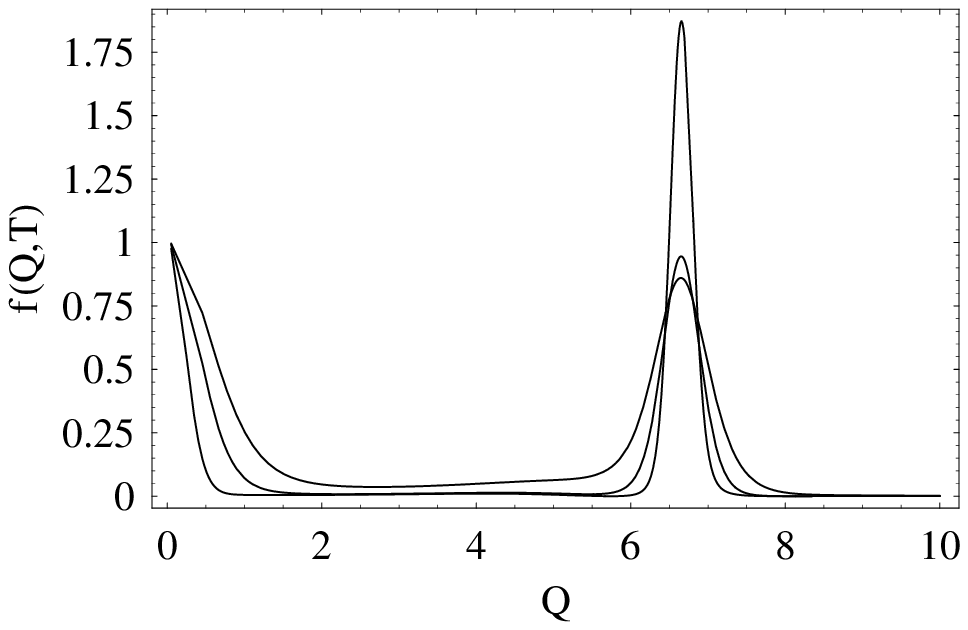}\label{fig:eta40b}}
\caption{Plots of $f(Q,T)$ for $\eta=0.40$ with (a) $R=0.52$ and (b) $R=0.56$. Times are (from top to bottom at $Q=0$) $T$=0.05, 0.1 and 0.3. Note that whereas the first structure peak of the $R=0.52$ plot decays to zero with time, the peak in the $R=0.56$ plot grows.}
\label{fig:eta40}
\end{figure}

\begin{figure}[btp]
\centering
\includegraphics[width=\columnwidth]{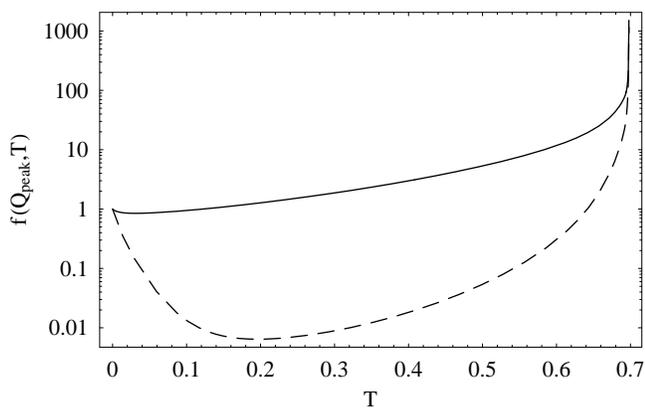}
\caption{Logarithmic plots of the first structure peak amplitude (solid line) and prepeak amplitude (dashed line) for $\eta=0.40$, $R=0.56$. Note that even though the first structure peak amplitude begins to grow first, the prepeak amplitude grows at a faster rate and the two are comparable when the solution breaks down.}
\label{fig:eta40amp}
\end{figure}

Moving to $\eta=0.40$, we again slowly increase the value of $R$. Initially, we clearly see the first structure peak and note that it decays to zero with time as in the $R=0$ case. We plot f(Q,T) for $R=0.52$ in Fig. \ref{fig:eta40a} as an example. Increasing $R$ a bit further to $R=0.56$ however, we cross over to a new regime. Now, the first structure peak amplitude initially decreases, then slows down and finally begins to increase. (Fig. \ref{fig:eta40b}.) If we continue the solution to long times, we find that this growth has no bound and in fact accelerates. The prepeak (which is initially only weakly visible) grows faster than the first structure peak and the two peaks are of comparable amplitude only after the model and the numerical solution break down. We plot the amplitude of the prepeak and first structure peak in Fig. \ref{fig:eta40amp}, including late times where the solution has become unphysical.

We may continue in this manner and look at higher values of $\eta$. We see a very general pattern appear. As $R$ is increased from zero, there is initially a period where all peaks decay to zero with time. The plots look qualitatively like the $R=0$ plots, however as $R$ is increased, the peaks decay more slowly. We call this region of parameter space the {\em stable} region. Continuing, we pass a critical value of $R$ and see that the amplitudes of the peaks no longer decay to zero, but turn around and grow without bound. We call this (interesting, yet unphysical) region the {\em unstable} region and label the critical coupling $R^*$. We find that as we increase $\eta$, the value of $R^*$ at which we cross from the stable to unstable regime decreases. As we continue increasing the value of $R$ past $R^*$, we find that the peak turns around its growth at earlier and earlier times.

The behavior of the prepeak is quite curious. As we increase $\eta$, we find that the prepeak becomes more difficult to see in the plots. Recall that at $\eta=0.30$, we could clearly see the prepeak even though we were in the stable region. At high values of $\eta$ we do not see the prepeak in the stable region and often do not see the prepeak in the unstable region until near the numerical breakdown when both the prepeak and first structure amplitudes become comparable. When both peaks are visible, we note that the first structure peak turns around and grows at an earlier time than the prepeak, but that the prepeak grows at a faster rate.

Though the prepeak is sometimes not visible from the raw data, we {\em can} extract some information about the prepeak's behavior. We believe that the behavior of the prepeak and the first structure peak, while quantitatively different, are related and that we do not lose information by monitoring just the latter. We discuss this in more detail in the sections which follow.

\section{Analysis}
\subsection{Stable-unstable crossover}
We have seen that our solution has two distinct regimes: the stable region and the unstable region. We now wish to quantify the crossover from stable to unstable behavior. Because this crossover depends on both the coupling constant and the density, we have a critical function $R^*(\eta)$.

\begin{figure}[btp]
\centering
\includegraphics[width=\columnwidth]{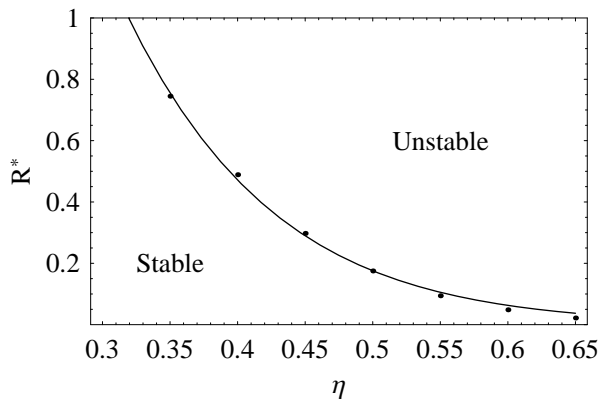}
\caption{Plot of the critical line $R^*(\eta)$ separating the stable and unstable regions. Each point was determined from a set of data using Eq. \eqref{eq:Tmin}. The solid line is to guide the eye only.}\label{fig:phase}
\end{figure}

To determine this line, we follow the approach developed in RDMI. Let us first hold $\eta$ fixed and vary $R$. If we are in the stable region, the amplitudes of the peak heights decrease monotonically with time. In the unstable region however, we may identify a time $T_{min}$ where the amplitude of a particular peak reaches its lowest point. (See, for example, Fig. \ref{fig:eta40amp}.) As we approach $R^*$ from larger R values, we find that the value of $T_{min}$ approaches infinity as we get arbitrarily close to $R^*$. We may therefore find $T_{min}$ for values of $R$ close to but larger than $R^*$ and fit these values to a function diverging at $R^*$:
\be
T_{min}=\frac{T_0}{(R-R^*)^x}.
\label{eq:Tmin}
\ee
If we repeat this technique for a set of $\eta$ values, we determine a corresponding set of $R^*$ values along the line $R^*(\eta)$ which divides the stable and unstable regimes \cite{Rstar}. Our results for $R^*(\eta)$ are plotted in Fig. \ref{fig:phase}.

With a technique in place, one now may ask which peak to use in determining $T_{min}$. We saw in the previous section that the prepeak and the first structure peak do not reach their minima at the same time and that their unstable growth is not equally paced. It is therefore unclear if the results from fitting one set of data will match the results from fitting the other. It is also unclear whether there may in fact be a region where we have unstable growth in one peak with stable decay in the other. A careful analysis, however, reveals that this is not the case. Values for $R^*$ determined from the prepeak amplitude closely match those determined from first structure peak amplitude and we are unable to find any region where the stability/instablilty of one peak does not match that of the other. Thus, we chose to use the first structure peak amplitude data for determining the critical function since it is the most clearly visible in all plots.

We can note several things from the plot in Fig. \ref{fig:phase}. First, we again see that for small to moderate values of $\eta$ the solution is stable for all studied values of $R$. The critical line does not cross the $R=1$ mark until some value between $\eta=0.30$ and $\eta=0.35$. Second, even though a transition to the unstable region at these moderate densities seems possible, one should be cautious about such a conclusion. Recall that we expect $R$ to be a small perturbation parameter. By this reasoning, we may wish to restrict ourselves to discussion of such a transition at higher densities where $R^*$ is indeed small. Such higher densities are are those more typical of very dense liquids, solids and glasses.

\subsection{Analytical Fits}

We now see that the dynamic structure factor has the general form
\be
f(Q,T)=f_0(Q,T)+f_{pp}(Q,T)+f_n(Q,T)
\ee
where $f_0(Q,T)$ is the initial peak centered at $Q=0$, $f_{pp}(Q,T)$ is the prepeak and $f_n(Q,T)$ are the peaks due to the structure factor, of which we are only concerned in the first ($n=1$). Each peak can be modeled as a Gaussian such that
\be
f_{x}(Q,T) = A_{x}(T)e^{-B_{x}(T)(Q-Q_{x})^2}
\label{eq:general}
\ee
where $x$ may be $0$, $pp$ or $1$ depending on the peak in question.

The peak centered at $Q=0$ is the simplest to treat. For small $Q$, we may neglect the interaction term and we are left with the de Gennes form given earlier by Eq. \eqref{eq:R0}:
\be
f(Q,T)=exp\bigg(\frac{-Q^2T\pi}{6\eta S(Q)}\bigg).
\ee
Comparing this to the generic Gaussian form of Eq. \eqref{eq:general}, we have $A_0 = 1$, $Q_0=0$ and
\be
B_0 = \frac{T \pi}{6\eta S(0)}.
\label{eq:B0}
\ee
A fit to the data shows excellent agreement. We find, for example, that for $\eta=0.55$, $R=0.14$, Eq. \eqref{eq:B0} predicts $B_0=102.38T$ while a fit yields $B_0=102.26T$.

\begin{figure}[btp]
\centering
\subfigure[]{\includegraphics[width=.48\columnwidth]{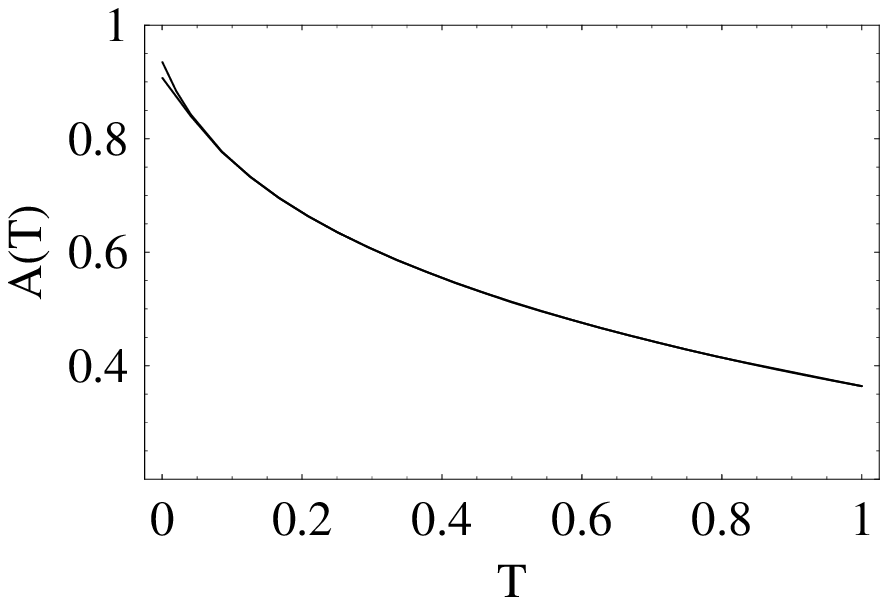}}
\subfigure[]{\includegraphics[width=.48\columnwidth]{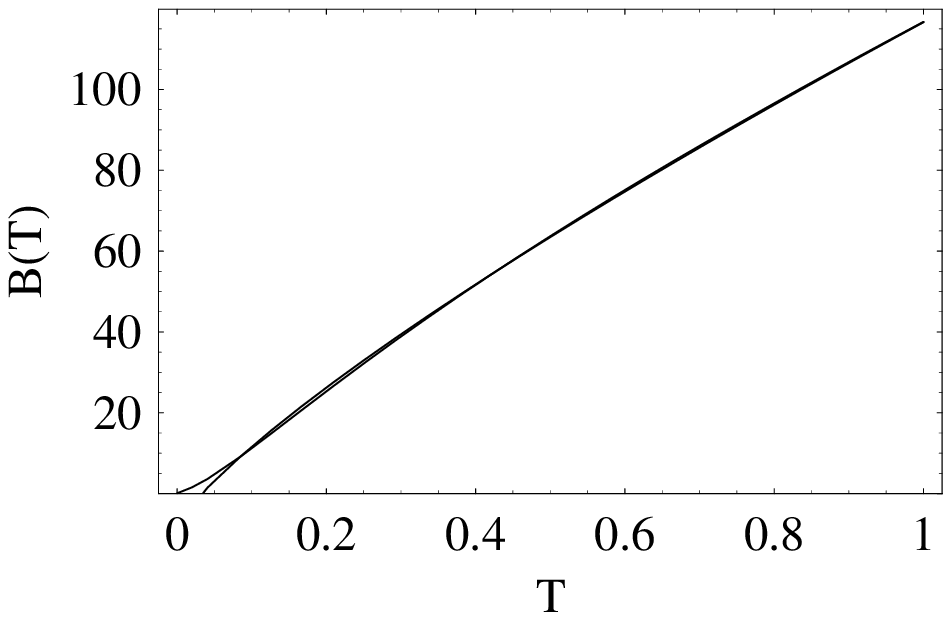}}
\subfigure[]{\includegraphics[width=\columnwidth]{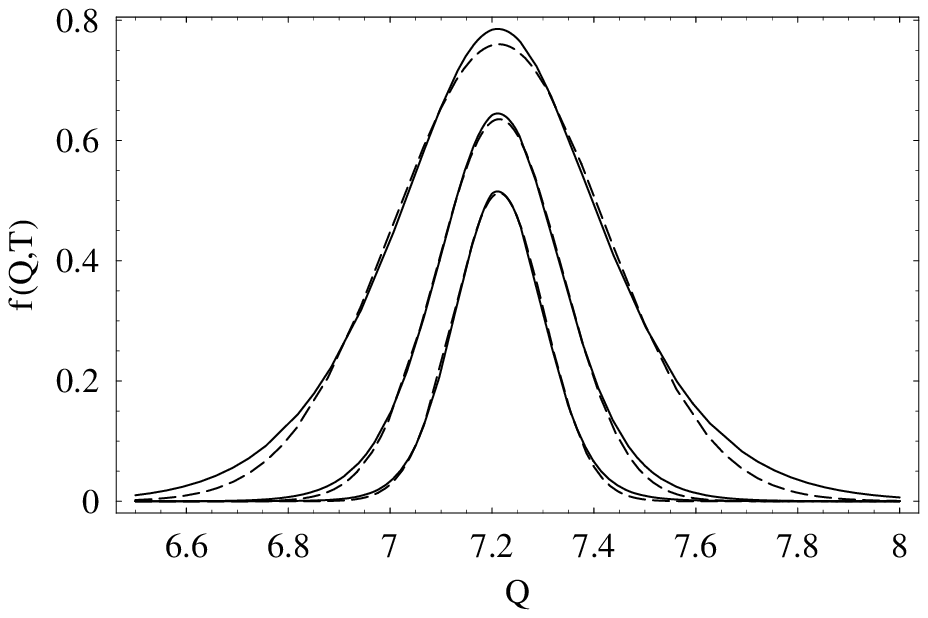}}
\caption{Plots of the (a) amplitude and (b) width fit to Eqs. \eqref{eq:A_x} and \eqref{eq:B_x} for $\eta=0.55$, $R=0.14$. In (a) the fit is given by $A(T)=0.593 e^{-0.478 T}/(T+0.0586)^{0.161}$ and in (b) by $B(T)=124.77 T^{0.801}-8.13$. The data and fit are indistinguishable except at the very earliest times. In (c), the structure peaks are shown (solid lines) with their Gaussian fits (dashed lines) at times $T= 0.1$, 0.25 and 0.5 (from largest to smallest amplitude). Note that the fit improves with time.}\label{fig:Fit_A_Fit_B}
\end{figure}

Near the prepeak and first structure peak, the situation is more complicated. In the stable regime, each Gaussian peak has an amplitude which fits nicely to
\be
A_x(T)=A_0 \frac{e^{-ET}}{(T+T_0)^{\alpha}}
\label{eq:A_x}
\ee
where the fit parameters vary with the coupling constant $R$. In the $R\rightarrow 0$ limit, we know from Eq. \eqref{eq:R0} that the first structure peak decays exponentially ($A_0=1$, $\alpha=0$, $E=T\pi/6\eta S(Q_1)$) and that the prepeak is non-existant ($A_0=0$). In the $R\rightarrow R^*$ limit, we find that both peaks tend toward pure power law decays ($E\rightarrow 0$). The peak width can be nicely fit to
\be
B_x=B_0 (T+T_0)^{\beta}
\label{eq:B_x}
\ee
where $\beta$ is approximately $1$ below the critical coupling, but decreases very rapidly as $R\rightarrow R^*$. The peak centers $Q_x$ drift slightly at early times, but quickly reach a fixed asymptotic value. Example fits for $A$ and $B$ can be seen in Fig. \ref{fig:Fit_A_Fit_B}.

In the unstable regime, peak amplitudes initially decrease then increase without bound and the peak widths narrow. At early times and away from $T_{min}$, the amplitudes may again be fit by Eq. \eqref{eq:A_x} and the widths by Eq. \eqref{eq:B_x}. However, the subsequent transition and growth pieces do not easily lend themselves to a universal fit, especially at late times when the growth accelerates and the numerical solution breaks down.

We find that these Gaussian peaks (in both the stable and unstable regimes) are tending toward delta functions as $T\rightarrow\infty$; the peak widths are narrowing to zero and the amplitudes are time-dependent. This is the same behavior which was seen with the prepeak in RDMI.

\subsection{Unstable Regime}
The behavior of the model in the unstable regime is clearly unphysical. What is happening? At first, one might suspect that the system is undergoing a liquid to glass transition. However, as was shown in RDMI, an ergodic non-ergodic transition can be found for this model at lowest order, but is not sustained at higher order. It is therefore more likely that the system is instead simply freezing from a liquid to a solid. Since the model is for a system in equilibrium, as the density and coupling are increased, the system may wish to nucleate. Our model, however, is not equipped to handle such a transition and so the peaks grow and are not properly stabilized.

One solution would be to extend the model by choosing an effective Hamiltonian known to support freezing. According to density functional theory, we can choose the effective Hamiltonian to be
\be
\beta {\cal H}_{\phi} = \int d^d x_1\bigg[\phi({\bf x_1}) \ln\bigg(\frac{\phi({\bf x_1})}{n}\bigg)
- \delta \phi({\bf x_1})\bigg]
\nonumber\\
-\frac{1}{2}\int d^d x_1 d^d x_2  \delta \phi({\bf x_1}) C_D({\bf x_1}-{\bf x_2})
\delta \phi({\bf x_2})
\ee
which, after expanding the logarithm and rearranging, generates higher order terms compared to the Hamiltonian used in this work:
\be
{\cal H}_{\phi}=\frac{1}{2} \int d^d x_1 d^d x_2 \delta \phi({\bf x_1}) \tilde{\chi}^{-1}({\bf x_1}-{\bf x_2})
\delta \phi({\bf x_2})
\nonumber\\
+ v\int d^d x_1 (\delta \phi({\bf x_1}))^3
+ u\int d^d x_1 (\delta \phi({\bf x_1}))^4 + \ldots
\ee
where
\be
\tilde{\chi}^{-1}({\bf x})=\chi^{-1}({\bf x})-\frac{\delta({\bf x})-n}{n\beta}
\ee
and the higher order coefficients are given by $v=-1/6\beta n^2$ and $u=1/12\beta n^3$.

A full treatment of this model with the above static perturbations will be given elsewhere.

\subsection{Prepeak Behavior}

We have seen that the prepeak is not always visible near and beyond the transition from stable to unstable. Let us explore the behavior and significance of the prepeak in greater depth.

\subsubsection{Extraction of Prepeak Information}

The prepeak is not always easily identifiable from the plots, but is hidden by other features in the data (e.g., the tails of the Q=0 peak and/or the first structure peak) and is of relatively small amplitude. To disentangle the prepeak from the other behavior, recall that we know the $R=0$ solution exactly (Eq. \eqref{eq:R0}) and that the solution includes no prepeak and has a purely exponential decay for the structure factor peaks.

Let us {\em divide out} the $R=0$ contribution to $f(Q,T)$,
\be
h(Q,T)=\frac{f(Q,T)}{f(Q,T)|_{R=0}}.
\label{eq:divided}
\ee
Though $f(Q,T)$ depends on $R$ through an integro-differential equation and cannot be separated into purely $R=0$ and $R\neq0$ contributions in either an additive or multiplicative fashion, the method has the potential to yield {\em some} information since we have seen that the late time behavior is a simple sum of Gaussian peaks. More explicitly, we expect
\be
h(Q,T)\sim\frac{f_0{Q,T}+f_{pp}(Q,T)+f_1(Q,T)}{f_0(Q,T)+f_1^{\prime}(Q,T)}
\ee
where the prime simply designates that while the form for the peak is the same, the amplitudes and widths differ \cite{h_trickery}.

\begin{figure}[btp]
\centering
\subfigure[]{\includegraphics[width=\columnwidth]{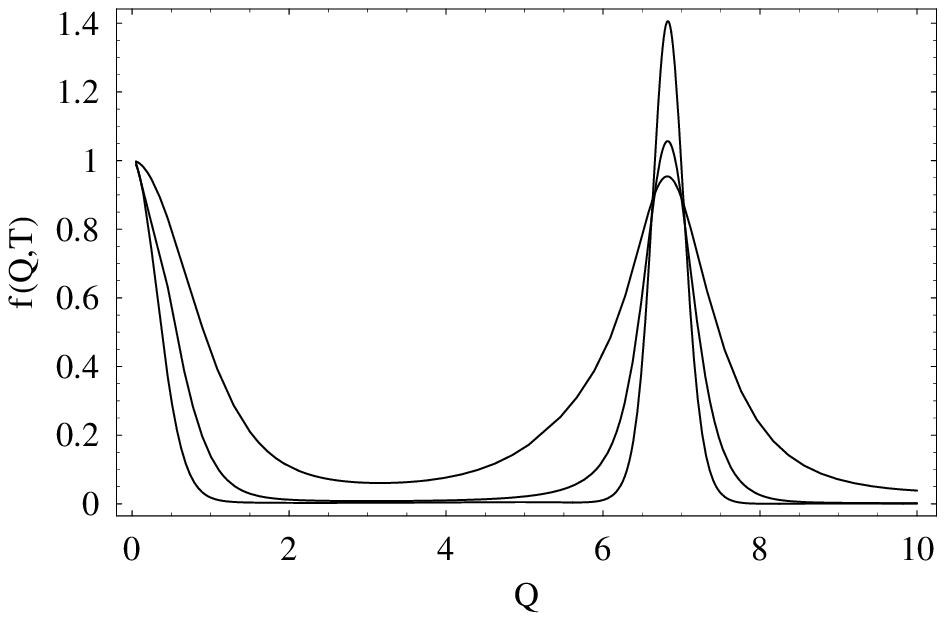}\label{fig:eta45raw}}
\subfigure[]{\includegraphics[width=\columnwidth]{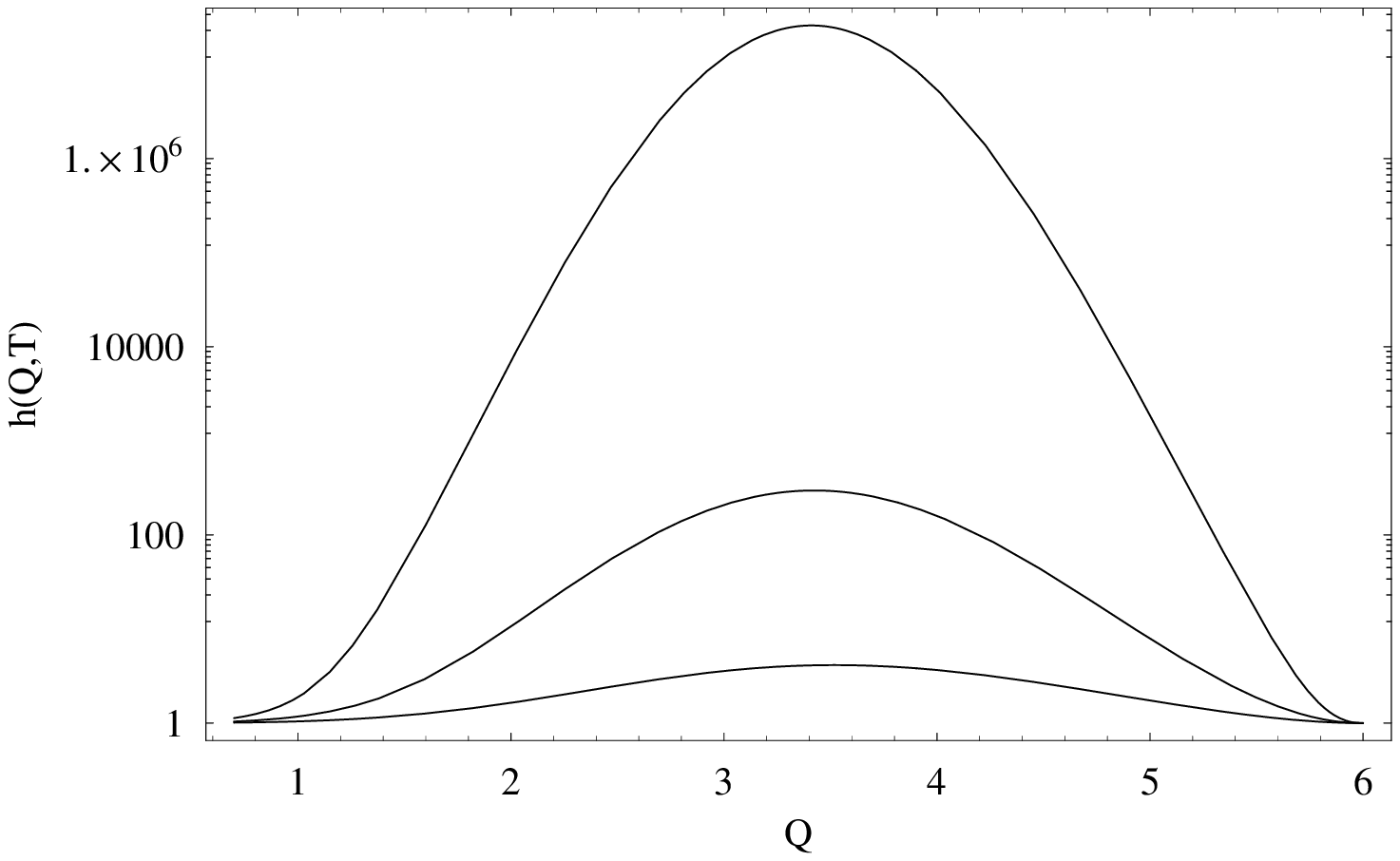}\label{fig:eta45div}}
\caption{(a) Plot of $f(Q,T)$ for $\eta=0.45$, $R=0.42$. (b) Logarithmic plot of $h(Q,T)$ (given by Eq. \eqref{eq:divided}) in the prepeak region at the same $\eta$ and $R$ values. Times are $T$=0.02, 0.05 and 0.1 (from smallest to largest peak amplitude). Note that while no prepeak is visible in f(Q,T), a rapidly growing Gaussian appears in h(Q,T)}
\end{figure}

As a test of the above proposed methods, let us look at the plot of $f(Q,T)$ for $\eta=0.45$, $R=0.42$ (Fig. \ref{fig:eta45raw}). We see that we are in the unstable region since the first structure peak is growing with time. However, we see no sign of any prepeak. Let us divide out the $\eta=0.45$, $R=0$ values from our original $f(Q,T)$ to get $h(Q,T)$ (Fig. \ref{fig:eta45div}). Now we find a very clear prepeak at $Q\approx3.40$ which grows very rapidly in amplitude and narrows in width with time along with a smaller peak in the location of the first structure peak which also grows and narrows.

If we repeat this technique with other combinations of $\eta$ and $R$ we find that regardless of the parameters, a prepeak centered near $Q\approx3.40$ can always be found. Also, whereas the prepeak and first structure peak centers exhibit some early time transient drift when the raw results are plotted, these new plots show fixed peak centers.

Let us now try to fit these peaks to Gaussians. The peak amplitudes can be fit to
\be
A(T)=A_0 (T+T_0)^{\alpha} e^{ET},
\ee
where $\alpha$ is positive for the first structure peak and negative for the prepeak. The width is given by
\be
B(T)=B_0 (T+T_0)^{\beta}
\ee
where $\beta$ is approximately $1$ for both peaks.

\subsubsection{Connection to Experimental and Simulation Results}

While a prepeak developed in both the structureless approximation of RDMI and our own more realistic calculations here, there is little experimental evidence for the existence of such a peak in hard-sphere liquids. For molecular systems, a number of experiments reveal prepeaks below the first structure peak \cite{Morineau,Comez1,Comez2}, but this work is likely unrelated to our model since the peaks can generally be explained through physical mechanisms unique to molecular systems (e.g. hydrogen bonding leading to clustering, molecular reconfigurations, etc.) and since these prepeaks generate \textit{static} structure prepeaks instead of the \textit{dynamic} structure we see in our model. Work with molecular dynamics simulations on the other hand, has shown some dynamically generated prepeaks \cite{Kaemmerer,Theis1,Theis2,Winkler}, but again this work is on molecular systems and can be explained by uniquely molecular dyanamics, e.g. translation-rotation coupling.

While no evidence for prepeaks in hard-sphere or monatomic systems has been reported, we show here that at higher densities the prepeak was not readily visible in the dynamic structure factor except in the (non-physical) long-time breakdown. It is therefore conceivable that a prepeak similar to that found here may in fact exist in experimental results, but has simply not been detected and extracted.

How could a prepeak hidden in experimental data be extracted? In the previous section, we showed that hidden prepeaks appear if one plots the raw data divided by the theoretically determined zero-coupling data as in Eq. \eqref{eq:divided}. This form, however, is not useful to an experimentalist, so let us rewrite it in terms of density $n$, temperature $\beta$ and the physical diffusion coefficient, $D_p={\bar D}$ -- parameters which would be known or could be determined -- and the static and dynamic structure factors ($S(q)$ and $C(q,t)$ respectively) -- which would be measured. We then have
\be
h(q,t) = \frac{C(q,t)}{nS(q)} \exp \bigg(\frac{q^2D_pt}{nS(q)\beta}\bigg).
\label{eq:exp}
\ee

For such a method to yield useful information, extremely good data resolution for the dynamic structure factor at small wavenumbers in the prepeak region would be needed, as would small uncertainties on the quantities going into Eq. \eqref{eq:exp}. Poor resolution and large uncertainties would easily wash out the small values that one is trying to extract.

\section{Conclusions}

We began with the random diffusion model introduced in RDMI which describes a system undergoing diffusive dynamics with a density-dependent bare diffusion coefficient. While the model had previously been studied in the structureless approximation where the short distance degrees of freedom had been integrated out, we wished to study the model from a more realistic view that included such features. By relaxing the delta function constraint of the transport matrix and introducing the smoothing functions $f_0(q)$ and $f_1(q)$, we were able to introduce a realistic structure factor back into the problem while still keeping the wavenumber integrals finite. The resulting kinetic equation then depended on two variable parameters, the dimensionless density $\eta$ and the coupling constant $R$.

In our investigations, we came to the following conclusions:

\begin{enumerate}
\item
When the coupling constant is increased, there is a significant slowing down of the system; exponential decay gives way to algebraic decay.

\item
For large enough coupling there is a transition where the system goes from stable to unstable. All peaks grow without bound and the solution becomes unphysical. The critical coupling required to reach this transition decreases as the density increases.

\item
It is possible this instability may be cut off by including terms in the model which allow for nucleation. We propose terms motivated by density functional theory and the results of this model extension will be pursued in another work.

\item
Near but below the transition, a new peak, termed the prepeak, sometimes can be seen between the $Q=0$ peak and the first structure peak. When it can be seen, its amplitude is less than that of the first structure peak. Above the transition, the prepeak always appears, though it may only be seen at late times when the peak amplitudes are growing without bound.

\item
Near but below the transition, both the prepeak and first structure peak narrow to delta functions, but with algebraically decaying amplitude.

\item
It is possible to isolate the prepeak by dividing out the $R=0$ solution. In such a case we find the prepeak and structure peaks to be well separated with growing amplitudes and narrowing widths. This technique can separate out the prepeak even when it cannot be seen in the raw data and we find that the prepeak is always centered near $Q\approx3.40$ regardless of coupling or density.

\item
By the above technique, one should be able to take an experimentally determined dynamic structure factor and isolate the prepeak using only experimentally measured or controlled parameters. Such an isolation might reveal previously undetected prepeak information and give indirect measurements of parameters of the system not usually accessible such as the the coefficients of the terms in the bare diffusion coefficient.

\end{enumerate}

These results confirm that the features seen in RDMI were not due to the nature of the structureless approximation, but were indeed inherent features of the model. In addition, the model is now sufficiently realistic to invite comparison with experimental results. While the prepeak seen in our model has not been reported experimentally, we find that at reasonably accessible parameter values, it is often hidden; we provide a method by which a hidden prepeak might be extracted. The model, however, still lacks a sufficient mechanism to allow for freezing. Future work introducing static perturbations to the Hamiltonian may rectify this.

\begin{acknowledgments}
This work was supported by the Joint Theory Institute and the Department of Physics at the University of Chicago.
\end{acknowledgments}

\appendix
\section{Method of Numerical Solution}
To solve the kinetic equation given by Eqs. \eqref{eq:final} and \eqref{eq:kernel}, we use a simple Euler-type integration scheme to find $f(Q,T)$ over a discrete set of wavenumbers $Q_i$ and times $T_j$. The $Q$ values are equally spaced, but the $T$ values are chosen using an adaptive step routine and stored in an array. In various integration steps, values of f(Q,T) at points off this lattice are sometimes required. Since f(Q,T) varies smoothly, a simple interpolation scheme is used to estimate such values. Let us address the integration method, the adaptive step routine and the interpolation scheme in turn. Methods were adapted from Ref. \onlinecite{NumRec}.

\subsection{Integration Method}
The kinetic equation (Eq. \eqref{eq:final})is presented in the simple form
\be
\frac{\partial f(Q,T)}{\partial T} = g(Q,T)
\ee
where
\be
g(Q,T)&=&-\frac{Q^2}{S(Q)} \frac{\pi}{6\eta}f(Q,T)
\nonumber\\
&&+R^2\int^T_0 dS N(Q,T-S)f(Q,S).
\ee
If we ignore the intricacies of $g(Q,T)$ for the moment, we can advance our solution to $f(Q,T)$  one time step at a time (in the normal Euler fashion) as
\be
f(Q_i,T_{j+1})&=&f(Q_i,T_j)
\nonumber\\
&&+(T_{j+1}-T_j)g(Q_i,T_j)
\ee
using the initial conditions that $f(Q_i,T_0)=1$.

The function $g(Q,T)$ includes a time integral over the memory kernel which can be decomposed into a similar differential equation,
\be
\frac{dh(Q,T,S)}{dS}=R^2N(Q,T-S)f(Q,S),
\ee
and can thus similarly be solved. Note now that since the time steps are not even, interpolation may be needed to find values of functions off the lattice. This will be addressed in detail below.

At the next nested level, the memory function is itself an integral over wavenumber. In three dimensions, $d{\bf K}=K^2 dKd\sin\theta d\phi$. Integrating over $\phi$ and expressing dot products as $|{\bf Q}\cdot {\bf K}| = \sqrt{Q^2+K^2-2QK\cos\theta}$ reduces the integral to two dimensions, which are again solved by the Euler method using interpolations.

Thus, to summarize, as we advance $f(Q,T)$ from one time step to the next, we must first advance the memory function array by one time step, then perform the time integration over it to get the function $g(Q,T)$ and finally use $g(Q,T)$ to fully advance $f(Q,T)$.

\subsection{Adaptive Step Routine}
We desire an adaptive time step size since we expect our solutions to sometimes offer smooth decays to zero and other times offer rapid growth and eventual instability. Let us describe the routine.

When choosing step sizes, there is always a trade-off between maximizing computation speed (larger step sizes) and minimizing error (smaller step sizes). To quantify this trade off, let us imagine advancing a function $y(x)$ from initial value $y(x_0)$ up to $y(x_f)$ in two different ways. In the first method, let us take two steps of size $h_1=(x_f-x_0)/2$ and call the result $Y_1$. In the second method, let us instead take one large step such that $h_2=2h_1$ and call that result $Y_2$. The difference between these two results is then $\triangle Y =|Y_1-Y_2|$. If one now specifies a desired difference between these values $\triangle Y_0$ -- which in some sense quantifies the trade off between accuracy and speed described above -- then we can infer that the best choice for the next step is given by
\be
h_{new}= h_1\frac{\triangle Y_0}{\triangle Y}.
\label{eq:StepSize}
\ee
We see that if the actual difference is smaller than our desired difference, we can afford to increase the step size, whereas if the actual difference is larger, we decrease our step size.

In our numerical integration, we specify an initial time step and tolerance $\triangle Y_0$. After each time step, the program determines where in $Q$-space the fastest growth/decay is occurring (where one assumes $\triangle Y$ will be the largest) and then computes the two test values described above at that point; the program integrates up from $f(Q,T_{j-1})$ to $f(Q,T_{j+1})$ first with two steps of the current size and then again with one step at twice that. From these points, the next step is determined.

Two limiting actions are taken to prevent run-away changes from occurring. First, a safety factor of $0.9$ is added to the right hand side of Eq. \eqref{eq:StepSize}. This serves to underestimate the new step size so that the adjustments always err on the side of smaller error rather than faster speed. Second, the new step size is never allowed to be more than four times the previous step size. This prevents dramatic changes in step size.

The choices of these limiting values ($0.9$ and four times) as well as our choice for the tolerance $\triangle Y_0=10^{-6}$ were chosen through rough trial and error.

\subsection{Interpolation Algorithm}
Integrations over the memory kernel require sampling the function at times not on the previously computed $Q_i$, $T_j$ lattice. In such cases, a two-dimensional bilinear interpolation was used. If one wants to know the value of $f(Q,T)$ and knows the value at surrounding points $(Q_1,T_1)$, $(Q_1,T_2)$, $(Q_2,T_1)$ and $(Q_2,T_2)$ where $Q_1<Q<Q_2$ and $T_1<T<T_2$, then we can estimate the value of $f(Q,T)$ as
\be
f(Q,T)&=&(1-t)(1-u)f(Q_1,T_1)+t(1-u)f(Q_2,T_1)
\nonumber\\
&& + t u f(q_2,T_2)+(1-t)u f(Q_1,T_2)
\ee
where we define
\be
t\equiv(Q-Q_1)/(Q_2-Q_1)
\ee
and
\be
u\equiv(T-T_1)/(T_2-T_1).
\ee


\begin{thebibliography}{99}

\bibitem{RDMI} G. F. Mazenko, Phys. Rev. E {\bf 78}, 031123 (2008).

\bibitem{FA1} G. H. Fredrickson and H. C. Andersen, Phys. Rev. Lett. {\bf 53}, 1244 (1984).

\bibitem{FA2}
F. Ritort and P. Sollich, Adv. Phys. {\bf 52}, 219 (2003).

\bibitem{chan1}
J. P. Garrahan and D. Chandler, PNAS {\bf 100}, 9710 (2003).

\bibitem{chan2}
J. P. Garrahan and D. Chandler, Phys. Rev. Lett. {\bf 89}, 035704 (2002).

\bibitem{WBG1}  S. Whitelam, L. Berthier, and J. P. Garrahan,
Phys. Rev. Lett. {\bf 92}, 185705 (2004).

\bibitem{WBG2} S. Whitelam, L. Berthier, and J. P. Garrahan,
Phys. Rev. E {\bf 71}, 026128 (2005).

\bibitem{jack} R. Jack, P. Mayer and P. Sollich, cond-mat/0601529.

\bibitem{Dean} D. S. Dean, J. Phys. A {\bf 29}, L613 (1996).

\bibitem{kaw}
K. Kawasaki and S. Miyazima, Z. Phys. B {\bf 103}, 423 (1997).

\bibitem{RFT} K. Miyasaki and D. R. Reichman, J. Phys. A {\bf 38}, L343 (2005).

\bibitem{TM}
We have some freedom in picking the form of this matrix, but it must be chosen to satisfy
\be
\int ~d^{d}x_{1}\frac{\delta \Gamma_{\phi}({\bf x}_{1},{\bf x}_{2})} {\delta \phi ({\bf x}_{1})}=0.
\nonumber
\ee
Eq. \eqref{eq:TM} is the most general polynomial from we have been able to construct which is compatible with this constraint.

\bibitem{cutoff}
This choice corresponds roughly to a short wavenumber approximation in the extended hydrodynamical limit. This correspondence can be seen by taking a static susceptibility which decays smoothly to zero for large-q given by
\be
\chi(q)=\chi_0 e^{-(q\ell)^2/2}
\nonumber
\ee
and looking at the behavior in the the small-q regime. If one associates $\Lambda \rightarrow 1/\ell$ and $r \rightarrow \chi_0^{-1}$, the memory function results are similar. For more details, see RDMI Section VI-C.

\bibitem{NESM}
A derivation of the kinetic equation in the Fokker-Planck formalism is given in Chapter 9 of G. Mazenko, {\bf Nonequilibrium Statistical Mechanics}, Wiley (Berlin), 2006.

\bibitem{Das}
S. P. Das, Rev. Mod. Phys. {\bf 76}, 786 (2004).

\bibitem{Goetze}
W. G\"{o}tze, {\bf Complex Dynamics of Glass-Forming Liquids: A Mode-Coupling Theory}, Oxford University Press, 2009.

\bibitem{ABL}
A. Andreanov, G. Biroli and A. Lef\`{e}vre, J. Stat. Mech. P07008 (2006)

\bibitem{KK}
K. Kawasaki and T. Koga, Physica A {\bf 201}, 115 (1993).

\bibitem{ashcroft}
N. W. Ashcroft and J. Lekner, Phys. Rev. {\bf 145}, 83 (1966).

\bibitem{Hansen}
J-P. Hansen and I.R. McDonald, {\bf Theory of Simple Liquids, Third Edition}, Academic Press, 2006.

\bibitem{Rstar}
In addition to $R^*$, the fit also yields the parameters $T_0$ and $x$. One can ask how these parameters vary as functions of $\eta$. Unfortunately, our numerical integration scheme does not allow us to probe regions arbitrarily close to the critical value due to finite memory restrictions. In approaching the critical point, we were able to generate reliable and consistent values for $R^*$ with each new data point, however the other parameters were less well behaved and varied considerably depending on which points were used in the fits. In general, $T_0$ decreases as $\eta$ increases while the $x$ values cluster about a central value indicating either a weak or nonesistent dependence on $\eta$. To get a better understanding of the behavior of these parameters would require an improved numerical integration scheme.

\bibitem{h_trickery}
Note that it is possible to divide one function without an extremum at a certain point by another without an extremum at that same point and generate a result \textit{with} an extremum. Care was taken to check that this is not what is occurring here. One can look at the raw data points for a given time and, given the amplitude and widths of each of the Gaussians in the simple model
\be
h(Q,T)\sim\frac{f_0{Q,T}+f_{pp}(Q,T)+f_1(Q,T)}{f_0(Q,T)+f_1^{\prime}(Q,T)}
\nonumber
\ee
show that if the amplitude of $f_{pp}$ is zero, no peak in $h(Q,T)$ will develop, but if the amplitude is greater than some very small threshold, a growing Gaussian peak develops as is seen in the numerical data.

\bibitem{Morineau}
D. Morineau, C. Alba-Simionesco and M.-C. Bellisent-Funel, Euro-phys. Lett. {\bf 43}, 195 (1998).

\bibitem{Comez1}
L. Comez, S. Corezzi, G. Monaco, R. Verbeni and D. Fioretto, Journal of Non-Crystalline Solids {\bf 352}, 4531-4535 (2006).

\bibitem{Comez2}
L. Comez, S. Corezzi, G. Monaco, R. Verbeni and D. Fioretto, Phys. Rev. Lett. {\bf 94}, 155702 (2005).

\bibitem{Kaemmerer}
S. K\"{a}mmerer, W. Kob and R. Schilling, Phys. Rev. E. {\bf 58}, 2131 (1998).

\bibitem{Theis1}
C. Theis, F. Sciortino, A. Latz, R. Schilling and P. Tartaglia, Phys. Rev. E. {\bf 62}, 1856 (2000).

\bibitem{Theis2}
C. Theis and R. Schilling, Journal of Non-Crystalline Solids {\bf 235-237}, 106 (1998).

\bibitem{Winkler}
A. Winkler, A. Latz, R. Schilling and C. Theis, Phys. Rev. E {\bf 62}, 8004 (2000).

\bibitem{NumRec}
W. H. Press, S. A. Teukolsky, W. T. Vetterling and B. P. Flannery, {\bf Numerical Recipes: The Art of Scientific Computing}, Cambridge University Press.

\end{thebibliography}
\end{document}